\documentclass[12pt]{article}
\usepackage{a4}
\usepackage{epsf}\usepackage{psfig}
\newcommand{\be}{\begin{equation}}
\newcommand{\ee}{\end{equation}}
\newcommand{\bd}{\begin{displaymath}}
\newcommand{\ed}{\end{displaymath}}
\newcommand{\bea}{\begin{eqnarray}}
\newcommand{\eea}{\end{eqnarray}}
\newcommand{\beao}{\begin{eqnarray*}}
\newcommand{\eeao}{\end{eqnarray*}}
\newcommand{\nn}{\nonumber}

\newcommand{\e}{{\rm e}} 
 
\renewcommand{\d}{{\rm d}}

\newcommand{\Ref}[1]{(\ref{#1})}
\newcommand{\ep}{{\epsilon}}

\newcommand{\E}{E}

\begin{document}
\title{Ground State Energy of Massive Scalar Field in the Background of
Finite Thickness Cosmic String}
\author{{\sc N.R. Khusnutdinov}${}^{1)}$\thanks{e-mail:
nail@dtp.ksu.ras.ru} {\sc \  and M. Bordag}${}^{2)}$\thanks{e-mail:
Michael.Bordag@itp.uni-leipzig.de}\\ \\ 
{\small 1)}{ Department of Physics, Kazan State Pedagogical University,}\\
Mezhlauk 1, 420021, Kazan, Russia, and \\
Engelhardt Observatory of Kazan, 422526, EAO, Russia\\
{\small 2)} Universit\"at Leipzig,  
Institut f\"ur Theoretische Physik\\
Augustusplatz 10/11, 04109 Leipzig, Germany} 
\maketitle
\begin{abstract}
We calculate the ground state energy of a massive scalar field in the
background of a cosmic string of finite thickness (Gott-Hiscock
metric). Using zeta functional regularization we discuss the
renormalization and the relevant heat kernel coefficients in
detail. The finite (non local) part of the ground state energy is
calculated in 2+1 dimensions in the approximation of a small mass
density of the string. By a numerical calculation it is shown to
vanish as a function of the radius of the string and of the parameter
$\xi$ of the nonconformal coupling.
\end{abstract}
\section{Introduction}\label{Sec1}
The Universe may have undergone a number of phase transitions since
the big bang due to the spontaneous symmetry breaking in gauge
theories. A number of topological objects may have been produced during
the expansion of the Universe \cite{Kibble} amongst which the cosmic strings
seem to be of particular interest as seeds for the density fluctuations
in the early Universe that are responsible for the formation of
galaxies (see for example \cite{HindmarchKibble}). Also, the gravitational
radiation produced by the formation of cosmic strings is part of cosmological
scenarios. 

Several models of strings have been suggested. First of all Vilenkin
has investigated the case of an infinitely thin cosmic string
\cite{Vilenkin}. The energy momentum tensor of this string has a delta
function like singular form.  The space-time is locally flat except in
the origin where it has a delta-shaped Riemann tensor
\cite{Doclady}. The vacuum expectation value of the stress-energy
tensor for different kinds of fields in that background has been
calculated both for zero temperature \cite{ZeroT} and non-zero
temperature \cite{NonZeroT} cases.

The vanishing thickness of the string causes known problems. The
vacuum expectation value of the stress - energy tensor has a non -
integrabile singularity in the origin which can be seen already from
dimensional considerations.  As a consequence, for the renormalization
of the ground state energy of quantum fields an additional counterterm
is required. It is known as the topological Kac term \cite{Kac}. This
additional part may be recognized as due to boundary condition at the
origin \cite{NonZeroT}.

The mentioned problems can be avoided by considering a string with finite
thickness. The simplest case is that of a constant matter density inside the
string. It has been considered in Refs. \cite{Gott,Hiscock}. The pressure $p$
and energy density ${\cal E}$ inside the string obey the condition $p + {\cal
  E} = 0$.  The exterior of this string is a conical space-time and the
interior is a constant curvature space (''cup'' space). The metric is smoothly
matched on the surface of the string but the scalar curvature has a laps on
it. In fact, this space is a cone with a smoothed origin. There is no
gravitational field outside the string in both above cases in opposite to the
Newtonian logarithmic gravitational potential of a thread-like matter
distribution. Note that this statement remains valid for an arbitrary radial
matter distribution inside the string, provided that the translational
invariance along the string is not broken.

The purpose of this paper is to calculate the ground state energy of a
massive scalar field in the background of a finite thickness cosmic
string.

In fact, we consider the $(2+1)$ dimensional case.  In zetafunctional
regularization, the ground state energy of a scalar field $\Phi$ is
given by
\be\label{gze}
 E_{0}=M^{2s}\frac 1 2 \zeta(s-\frac 1 2 ),\ee
where
\be \label{zeta}
\zeta(s)= \sum\limits_{(n)}(\lambda_{(n)} + m^{2})^{-s}\ee
is the zetafunction of the corresponding Laplace operator. The parameter $M$
is arbitrary. It has the dimension of a mass. Usually it is denoted by $\mu$
which we reserve here for the linear mass density of the string.  We assume
the field $\Phi$ to be put into a large volume with Dirichlet boundary
conditions in order to render the eigenvalues discrete. It will be seen that
the influence of this boundary separates completely from the contributions of
the pure background.  The $\lambda_{(n)}$ are eigenvalues of the two
dimensional Laplace operator
\be \label{ev} 
(\Delta-\xi {\rm R})\varphi_{(n)}(x)=\lambda_{(n)} \ \varphi_{(n)}(x).  
\ee
where ${\rm R}$ is the curvature scalar.

In the $(3+1)$ dimensional case we would have to add the integration over the
momentum of the translational invariant direction along the axis of the string
and get
\be \label{gze3}
E_{0}^{(3+1)}={\sqrt{\pi}\over 2}{\Gamma(s-1)\over \Gamma(s-1/2)}
M^{2s}\zeta(s-1).
\ee
The ultraviolet divergencies of the ground state energy are completely
determined by the first few heat kernel coefficients. By means of
\be \label{zetak}
\zeta(s)=\int\limits_{0}^{\infty}{\d t\over t}{t^{s}\over
  \Gamma(s)}K(t)
 \ee
and the asymptotic expansion for $t\to 0$
\be \label{k}
K(t)={1\over 4\pi t}\sum\limits_{n\ge 0} B_{n}t^{n}
\ee
of the heat kernel $K(t)$ corresponding to the operator in
Eq. \Ref{ev} the divergent part of the ground state energy can be
expressed in terms of the first four coefficients (in (3+1) dimensions
five coefficients would enter). We define
\bea \label{ehk}
E_{0}^{\rm div}(s)&=&\left(\frac{M}{m}\right)^{2s}{1\over 8\pi}\left\{
\frac{\Gamma (s-3/2)}{\Gamma (s-1/2)} B_0 m^3 + B_{1/2} \frac{\Gamma
(s-1) m^{2}}{\Gamma (s - 1/2)} + B_1 m \right.\nonumber \\
&&+ \left.B_{3/2} \frac{\Gamma
(s)}{\Gamma (s - 1/2)} \right\}. 
\eea 
In the following we calculate these coefficients for a string of zero
thickness, reobtaining known results, and for a string of finite
thickness in the approximation of a small mass density of the
string. Using these coefficients, the renormalization of the ground
state energy can be carried out in the standard way by adding the
counterterms corresponding to \Ref{ehk} to a suitably defined
classical energy. So we get the renormalized ground state energy
\be \label{eren}
E_{0}^{\rm ren}=\lim_{s\to 0}\left(E_{0}-E_{0}^{\rm
    div}\right).
\ee
In the $(2+1)$ dimensional case we obtain the result $E_{0}^{\rm ren}=0$ in
the given order of small mass density numerically.

We note that the ground state energy defined in this way obeys the
normalization condition
\be\label{normcond} \E_{0}^{\rm ren}\to 0~~\mbox{for}~~m\to\infty\,,  
\ee
which follows, from the circumstance that the heat kernel expansion is
at once the asymptotic expansion for large mass.

The organization of the papers as follows. In Sec.\ref{Sec2} we
describe the Gott-Hiscock space-time of finite thickness cosmic
string. In Sec.\ref{Sec3} we write down the general formulas for the
thin string and calculate the corresponding heat kernel
coefficients. In the next section we do the same for the finite
thickness string. In Sec.\ref{Sec5} we calculate the ground state
energy in the approximation of a small angle deficit. The result is
discussed in Sec.\ref{Sec6} and an appendix contains some technical
details of numerical calculations.

We use units $\hbar=c=G=1$.
\section{The Space-Time}\label{Sec2}

\begin{figure}
             \epsfxsize=6truecm
\centerline{\epsfbox{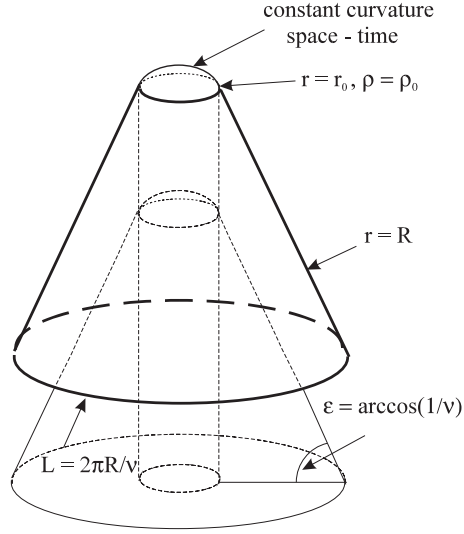} } 
             \vspace{.5cm}
\caption{Plot of two-dimensional $(t=const, z=const)$ part of finite
thickness cosmic string space-time}\label{Cone}
             \end{figure}

The metric of Gott - Hiscock \cite{Gott,Hiscock} is a solution of
Einstein's equations and it describes the space-time of an infinitely
long straight cosmic string with non vanishing thickness. The energy
density $\cal{E}$ is constant inside the string and it is zero outside
of it. The manifold can be covered by two maps. The first map, $t\in
(-\infty , +\infty ) , \rho \in [0, \rho_0] , \varphi \in [0, 2\pi ]
,z \in (-\infty , +\infty )$, covers the interior of the string and
the second one, $t\in (-\infty , +\infty ) , r \in [r_0, +\infty ) ,
\varphi \in [0, 2\pi ] ,z \in (-\infty , +\infty ) $, covers the
exterior. The coordinates $(t, \varphi, z)$ are the same in both
maps. $r_0$ and $\rho_0$ denote the radius of the string in external
and internal coordinates, respectively. The string is situated along
the $z$-axis. The metrics are $C^1$ -- matched on the surface of the
string, there is no surface stress energy (the extrinsic curvature
tensors of the interior and exterior metrics are equal to each other
\cite{Hiscock}). The metric has the following form
\be
ds_{in}^2=dt^2-d\rho^2-\rho_*^2\sin^2(\frac{\rho}{\rho_*})
d\varphi^2-dz^2\ ,
\label{m1}
\ee 
inside the string, and 
\be
ds_{out}^2=dt^2-dr^2-\frac{r^2}{\nu^2}d\varphi^2-dz^2\ ,
\label{m2}
\ee 
outside of it. Here $\rho_*=1/\sqrt{8\pi\cal{E}}$ is 'energetic' radius of
the string; $\cal{E}$ is the energy density inside the string.  The matching
condition on the surface of the string links the exterior parameters $( \nu ,
r_0 )$ and interior ones $( \rho_* , \rho_0 )$ of the string
%
\be
\frac{\rho_0}{\rho_*} \stackrel{\rm def}{=} \epsilon = const\ ,\ \nu
=\frac{1}{\cos\epsilon} \ ,\ \frac{r_0}{\rho_0} =
\frac{\tan\epsilon}{\epsilon}\ . \label{condition}
\ee
From these relations we have the following consequences.  The limit to
the Minkowski space-time is achieved by letting the energy density
inside the string tend to zero $(\rho_* \to \infty )$ for fixed radius
of the string $\rho_0$.  Then the angle deficit will tend to zero too
because of $\epsilon \to 0$.  Thereby in this limit $\nu = 1/\cos
\epsilon = 1,\ r_0 = \rho_0$ and both metrics turn into Minkowski
space time. On the other hand side, in order to shrink the string
$(\rho_0 \to 0)$ at fixed exterior ($\epsilon$ resp. the angle
deficite are constant) we must turn the energy density ${\cal E}$ to
infinity proportional to $\epsilon^2/8\pi\rho_0^2$. Nevertheless the
energy $\mu$ per unit length of the string $\mu$ which is the product
of the energy density $\epsilon^2/8\pi\rho_0^2$ and the cross section
of the string is always constant and equals $(1 - 1/\nu)/4$, the same
value as for the infinitely thin cosmic string and it doesn't depend
on the radius of the string. The two dimensional part $(t=const, z=
const)$ of the space-time (\ref{m1}, \ref{m2}) is depicted in
Fig.\ref{Cone}.

The manifold can be covered also by one map. One can continue the
exterior radial coordinate $r$ into the interior of the string by
mapping $r=r_0 + ( \rho_0/\epsilon )\tan (\epsilon\rho /\rho_0 -
\epsilon )$ and the space-time will be described by the metric 
\beao
ds_{in}^2 &=& dt^2 - \frac{dr^2}{[1 + \epsilon^2(r-r_0)^2/\rho_0^2]^2} -
\frac{r^2 }{\nu^2}\frac{d\varphi^2}{1 + \epsilon^2(r-r_0)^2/\rho_0^2}
- dz^2,\ r \in [0,r_0]\ , \\ 
ds_{out}^2 &=& dt^2 - dr^2 - \frac{r^2 }{\nu^2}d\varphi^2 - dz^2\ ,\
r\in [r_0, \infty )\ .   
\eeao
Here, the parameters $\nu , \rho_0$ and $r_0$ are connected by condition
(\ref{condition}). Nevertheless we shall use the metric in two maps
because it is simpler for calculations. As far as the angle deficit
is fixed, that is $\epsilon = \arccos 1/\nu = \rho_0/\rho_*$ is
constant, one can exclude $\rho_*$ and rewrite the metric in the
form which will be used in the following 
\be\label{m1_1} 
ds_{in}^2=dt^2-d\rho^2-\frac{\rho_0^2}{\epsilon^2} \sin^2(\frac{\epsilon
  \rho}{\rho_0})d\varphi^2-dz^2\ ,\ \rho\in [0,\rho_0]\ ,  
\ee
inside the string, and   
\be\label{m2_1} 
ds_{out}^2=dt^2-dr^2-\frac{r^2}{\nu^2}d\varphi^2-dz^2\ ,\ r\in [r_0, \infty
)\,  
\ee
 outside of it.

\section{Infinitely thin cosmic string}\label{Sec3}
The metric for an infinitely thin cosmic string is given by \Ref{m2_1} for
$r\in[0,\infty)$. In the (2+1) dimensional case, which we consider here, the
coordinate $z$ is absent.  The eigenvalues $\lambda_{(n)}$ which enter the
zeta function \Ref{zeta} are determined by Eq. \Ref{ev} in the background of
the spatial part of this metric. The curvature scalar ${\rm R}$ in this
equation is proportional to the two dimensional delta function. So it is a
  potential with pointlike support and should be taken into account by
a self adjoint extension of the operator corresponding to Eq. \Ref{ev}.
Instead, we drop these contributions here by considering the case
$\xi= 0$ only.  

By means of the ansatz
\be \label{ansatz}
\varphi_{(n)} = e^{i n \varphi }\sqrt{\frac{\nu}{r}} {\cal R}(r)\ ,
\ee
$(n = 0, \pm 1, \pm 2 , \dots)$ we arrive at the radial equation
\be\label{radeq} \left( \frac{d^2}{d r^2} - \frac{n^2 \nu^2 -
1/4}{r^2} + \lambda^2 \right) {\cal R} = 0\ .  
\ee
 The solution regular at the origin of this equation is a Bessel
function
\be\label{regsol}
{\cal R} = \sqrt{\frac{\pi \lambda r}{2}} J_{n \nu} (\lambda r)\ .
\ee
We assume it to obey Dirichlet boundary conditions at $r=R$. Then the
solutions $\lambda=\lambda_{n,i}$ ($i=1,2,\dots$) of the equation
\be \label{radcond}
\sqrt{\frac{\pi \lambda R}{2}} J_{n \nu} (\lambda R) =0\ 
\ee
are the discrete eigenvalues $\lambda\to\lambda_{(n)} =
\lambda_{n,i}$.  Now the ground state energy and equally the zeta
function \Ref{zeta} can be written in the form
\be \label{zeta1}
 \E_{0}^{\rm thin}(s)=\frac12 M^{2s}
\sum_{i=1}^{\infty}\sum_{n=-\infty}^{\infty}(\lambda_{n,i}+m^{2})^{1/2-s}.
\ee
The sum over $i$ can be rewritten as an integral (for details see \cite{bekl})
\be\label{zeta2}
\E_{0}^{\rm thin}(s)=-M^{2s}\frac{\cos (\pi s)}{2\pi}
\sum_{n=0}^{\infty} d_n \int_m^\infty d k\  (k^2 - m^2)^{1/2- s}
\frac{\partial}{\partial k} \ln k^{-n\nu} I_{n\nu}(k R)\ ,
\ee
where $d_{n>0} = 2$ and $d_0 = 1$ is the multiplicity of the angular momentum.
Note the factor $k^{-n\nu}$ in the argument of the logarithm, see \cite{bekl}.

In order to investigate the pole of $\E_{0}^{\rm thin}(s)$ we use the
uniform asymptotic expansion of the modified Bessel function for $n\to\infty$
\cite{Abramowitz}
\be\label{ufe}
I_{n\nu}(n\nu z) = \sqrt{\frac{t}{2\pi n\nu}} e^{n\nu \eta (z)}
\left\{ 1 + \sum_{k=1}^\infty \frac{u_k(t)}{(n \nu)^k} \right\}\   
\ee
with $t = 1/\sqrt{1 + z^2}\ ,\ \eta (z) = \sqrt{1 + z^2} + \ln (z/(1 + \sqrt{1
  + z^2}))$ and $z = kR/n\nu$.

The pole term of the zeta function $\zeta(s)$ \Ref{zeta} in $s=-1/2$
resp. of the energy $\E_{0}(s)$ in $s=0$ will be delivered by the
first few terms (up to $k=2$) of this expansion when inserting them
into $\E_{0}$ \Ref{zeta2}. We note that this expansion is at once an
asymptotic expansion for large masses and that the individual heat
kernel coefficients enter expansions like \Ref{ehk} multiplied by the
corresponding power of $m$. Therefore we will in the following keep
track of these powers and drop all contributions which for
$m\to\infty$ are of order $O(\frac1{m})$.

After inserting the expansion \Ref{ufe} into \Ref{zeta2}, the integration
over $k$ can be carried out using 
\bd
\int_1^\infty dx (x^2 - 1)^{1/2 -s} x (1 + x^2/\alpha^2 )^{-p/2} =
\frac{\Gamma (\frac{3}{2} - s) \Gamma (s + \frac{p - 3}{2})}{ 2\Gamma
(\frac{p}{2})} \alpha^p (1 + \alpha^2 )^{-s - \frac{p-3}{2}}\ ,
\ed
and we obtain the following expression for $E(s)$  
\bea
E(s)&=& -\frac{\cos \pi s}{4\pi }\left(\frac{M}{m}\right)^{2s}R m^{2}
\Gamma (\frac{3}{2} - s) \left\{ \sum_{n=0}^\infty d_n \left[
\frac{\Gamma (s - 1)}{\sqrt{\pi}} {}_2F_1 -\frac{n\nu}{b 
}\Gamma (s -\frac{1}{2}) \right]\right. \nonumber \\
&&- \frac{\Gamma (s - \frac{1}{2})}{2b  } Z(0,s -
\frac{1}{2}) - \frac{1}{4b ^2 \sqrt{\pi}} \left[ \Gamma
(s) Z(0,s) - \frac{10}{3} \frac{\Gamma (s + 1)}{b ^2 } Z(2, s + 1)
\right] \nonumber \\ 
&&- \frac{1}{8b ^3} \left[ \Gamma (s + \frac{1}{2}) Z(0, s
+ \frac{1}{2}) - \frac{6}{b ^2} \Gamma (s + \frac{3}{2}) Z(2, s
+ \frac{3}{2})\right. \nonumber \\ 
&&+ \left. \frac{5}{2b ^4} \Gamma (s + \frac{5}{2}) Z(4, s
+ \frac{5}{2})\right] - \frac{1}{96 b ^4 \sqrt{\pi}} \left[
25 \Gamma (s + 1) Z(0, s + 1)\phantom{\frac{1}{2}} \right. \nonumber \\
&&-\frac{1062}{5b ^2} \Gamma (s + 2)
Z(2, s + 2) + \frac{884}{5b ^4} \Gamma (s + 3) Z(4, s + 3) \nonumber \\
&&- \left.\left.
\frac{1768}{63 b ^6} \Gamma (s + 4) Z(6, s + 4) \right] + \dots \right\}\ .
\label{ZetaN} 
\eea
Here ${}_2F_1 = {}_2F_1 \left(-\frac{1}{2}, s - 1; \frac{1}{2}; -
\left( \frac{n\nu}{b  }\right)^2 \right)$ is the hypergeometric
function; $Z(p,q) = \sum_{n=0}^\infty d_n (n\nu)^p (1 +
(\frac{n\nu}{b })^2)^{-q}$ and $b = mR$. Next we have to perform the
analytical continuation $s\to 0$. 

First of all let's consider the part containing the hypergeometric function
\bea
Y(s)&=&\sum_{n=0}^\infty d_n \left[ \frac{\Gamma (s - 1)}{\sqrt{\pi}}
{}_2F_1 -\frac{n\nu}{b  }\Gamma (s -\frac{1}{2}) \right] \nonumber \\
&=& \frac{\Gamma (s - 1)}{\sqrt{\pi}} + 2 \sum_{n=1}^\infty
\left[ \frac{\Gamma (s - 1)}{\sqrt{\pi}} {}_2F_1 -\frac{n\nu}{b 
}\Gamma (s -\frac{1}{2}) \right]\ .
\label{F21}
\eea
For the calculation of the series we use the Mellin - Barnes type
representation of the hypergeometric function 
\beao
&&\frac{\Gamma (s - 1)}{\sqrt{\pi}}{}_2F_1
\left(-\frac{1}{2}, s - 1; \frac{1}{2}; - \left( \frac{n\nu}{ 
b}\right)^2 \right) \\ 
&=& \frac{1}{\Gamma (-\frac{1}{2})}\frac{1}{2\pi i}
\int_{\gamma } \frac{\Gamma (s - 1 + t)}{t - 1/2} \Gamma (-t) \left(
\frac{n\nu }{b  }\right)^{2t} dt \ ,
\eeao
where the contour is such that the poles of $\Gamma (s - 1 + t)/(t -
1/2)$ lie to the left of it and the poles of $\Gamma (-t)$ to the
right \cite{Abramowitz}. Before interchanging sum and integral one has
to shift the contour $\gamma$ to the left crossing the pole at $t =
1/2$ up to $t = - 1/2$ and then to close it to the left. This is
because of the convergence of the series $\sum_{n=1}^\infty n^{2t}$ it
is necessary to have $t < - 1/2$. The residue at the point $t = 1/2$
cancels the second manifestly divergent term in the sum
(\ref{F21}). Taking the limit $s \to 0$ we obtain the following finite
expression for (\ref{F21})
\beao
Y(0)&=&\frac{4}{\sqrt{\pi}}\left(\frac{\nu}{b  }\right)^2
\zeta'_R(-2) + \frac{1}{\sqrt{\pi}} \left(2\ln \frac{2\pi b }{\nu} -
3\right) \\ 
&&+ \frac{1}{\sqrt{\pi}} \sum_{n=0}^\infty 
\frac{(-1)^n}{(n+1)(n+2)} \frac{\zeta_R (2n+2)}{n+3/2} \left(
\frac{b }{ \nu}\right)^{2 + 2n}\ .
\eeao
Here $\zeta_R(a)$ is Riemann zeta function. 

The series is absolutely convergent for $b /\nu < 1$, but we need it
in the domain of large $R$, that is for $b \gg 1$.  For this reason we
perform the analytic continuation into the domain we need. With $t =
ib /\nu$ in the formula
\be \sum_{n=1}^\infty \frac{t^{2n}}{n}\zeta_R (2n) = \ln \frac{\pi
t}{\sin \pi t}\ ,\ |t| < 1\ ,
\label{ZetaSeries}
\ee 
we can express the series in $Y(0)$ in terms of the series
(\ref{ZetaSeries}).  As a result, $Y(0)$ is expressed in terms of
functions which are analytical in the whole plane of $b$ and it can be
divided into a polynomial and an exponentially small part
\bd
Y(0) = \frac{2\sqrt{\pi}}{3} \frac{b }{\nu} - \frac{\sqrt{\pi} \nu}{3
b } + \frac{2}{\sqrt{\pi}} \ln \left( 1 -
e^{-\frac{2\pi b }{\nu}}\right) - 2\sqrt{\pi}
\left[4\widetilde{Q}_1 (b  /\nu ) - 2 \widetilde{Q}_2 (b 
/\nu)\right]\ 
\ed
with 
\be\label{Qtilde}
\widetilde{Q}_{a} (x) = \frac{1}{x^{a}} \int_0^x \frac{dt
t^{a}}{e^{2\pi t} - 1} - \frac{1}{x^{a}}
\frac{\Gamma (a + 1)}{(2\pi)^{a + 1}}\zeta_R (a + 1) = -
\frac{1}{x^{a}} \int_x^\infty \frac{dt t^{a}}{e^{2\pi t} - 
1}\ .
\ee
For large $b$ this expression is exponentially small. Therefor we
arrive at
\bd
Y(0) = \frac{2\sqrt{\pi}}{3} \frac{b }{\nu} - \frac{\sqrt{\pi} \nu}{3
b } + O(e^{-b})\ .
\ed
The same result may be obtained in another way. One can use an
analytical continuation of the hypergeometric function
\cite{Abramowitz}, namely
\beao
&& {}_2F_1\left(-\frac{1}{2}, s-1; \frac{1}{2};
-\left(\frac{n\nu}{b}\right)^2 \right) 
     = 
\frac{n\nu}{b} \frac{\Gamma (1/2) \Gamma (s - 1/2)}{\Gamma (s-1)} \\
&+&
\frac{\Gamma (1/2) \Gamma (1/2 - s)}{\Gamma (-1/2) \Gamma (3/2 - s)}
\left( 1 + \left(\frac{n\nu}{b} \right)^2 \right)^{1-s}
{}_2F_1\left(1, s-1; s + \frac{1}{2}; \frac{1}{1 +
\left(\frac{n\nu}{b}\right)^2 }\right) \ . 
\eeao
The first term in the rhs. cancels the second, divergent term in the
sum (\ref{F21}).  Next, one can use power series expansion for the
hypergeometric function because its argument $1/(1 + (n\nu/b)^2)$ is
always smaller then unity. The result will be the same as that
obtained above by a longer calculation.

The series $Z(p,q)$ in (\ref{ZetaN}) can be expressed in terms of the
Epstein - Hurwitz zeta function \cite{ElizaldeBook}
\bd
\zeta_{EH} (r,c) = \sum_{n=1}^\infty (n^2 + c^2)^{-r}\ .
\ed
A known, quickly convergent expansion for large values of the
parameter $c$ is
\bd
\zeta_{EH}(r,c) = -\frac{c^{-2r}}{2} + \frac{\sqrt{\pi}\Gamma (r -
1/2)}{2\Gamma (r)} c^{-2r + 1} + \frac{2 \pi^r c^{-r + 1/2}}{\Gamma
(r)} \sum_{n=1}^\infty n^{r - 1/2} K_{r - 1/2} (2\pi n c)\ .
\ed
For small $c$ it holds
\bd
\zeta_{EH}(r,c)= \sum_{n=0}^\infty (-1)^n \frac{\Gamma (n +
r)}{\Gamma (r) n!} c^{2r} \zeta_R (2r + 2n)\ .
\ed
For integer $r \ge 1 $ the zeta function can be expressed in terms of
elementary functions. So we get the relevant part of the ground state
energy in the form
\bea\label{zetaexp}
\E_{0}^{\rm thin}(s) &=& \left(\frac{M}{m}\right)^{2s}\frac{1}{8\pi }
\left( -\frac{2\pi R^2}{3\nu} m^3 - \frac{\pi^{3/2} R m^2\Gamma (s
-1)}{\nu \Gamma (s - \frac{1}{2})} + \frac{\pi}{3}\left(\nu +
\frac{1}{\nu}\right) m \right. \nonumber \\
&&+ \left.\frac{\pi^{3/2} \Gamma (s)}{ 32 R\nu \Gamma (s -
\frac{1}{2})} \right) + \frac1R O\left(\frac{1}{Rm}\right) , 
\eea
dropping contributions proportional to $\exp (-Rm)$. By comparing
this formula with $\E_{0}^{\rm div}$ \Ref{ehk} we can read off 
the heat kernel coefficients
\be\label{hkks}
B_0 = \frac{\pi R^2}{\nu}\ ,\ B_{1/2} = - \frac{\pi^{3/2}R}{\nu}\ ,\
B_1 = \frac{\pi}{3}\left( \nu + \frac{1}{\nu}\right)\ ,\ B_{3/2} =
\frac{\pi^{3/2}}{32 R\nu}\  .
\ee
Now, taking into account the general structure 
\be
B_{r}= \int_{\partial V} c_r dS + \int_V a_r dV   
\ee
of the coefficients we represent $B_{1}$ as
\be\label{B-1}
B_1 = \frac{2\pi}{3\nu} + \frac{\pi}{3} \left( \nu -
\frac{1}{\nu}\right)\ . 
\ee
In this representation, the first term in $B_{1}$ and all other
coefficients are seen to be the result of the boundary at $r=R$. In
fact, they are known \cite{Dowker}. The second term in $B_{1}$ is
independent on that boundary. It is known as the topological Kac term
\cite{Kac} and is a result of the conical singularity.

Using these coefficients, by means of Eq.\Ref{ehk}, we can define
  $E_{0}^{\rm div}$ and the renormalized ground state energy
  $\E_{0}^{\rm ren}$ \Ref{eren}. Now we observe, that all
  contributions in \Ref{zetaexp} except for the topological Kac term
  are due the the boundary at $r=R$. Leaving them aside, only the Kac
  term remains. It must be included into $\E_{0}^{\rm div}$ and we get
  \be\label{Eren0} E_{0}^{\rm ren}=0 
\ee
for the genuine contributions of the string, or equally in the limit
$R\to\infty$. Let us remark, that this result holds also in (3+1) and
higher dimensions as can be inferred from dimensional reasons. The
parameters entering the problem are the mass density of the string
which enters together with the graviational constant to form a
dimensionless combination, expressed by the angle deficite, for
example and the mass of the quantum field. In the case of a string
with zero radius there is no further dimensional parameter on which
the renormalized ground state energy might depend. As the ground state
energy has the dimension of a inverse length it might be proportional
to the mass but such terms had been subtracted in the renormalization
process. Note that this discussion does not apply to the case of a
string with finite radius because this radius is the additional
dimensional parameter allowing in general for a nontrivial
renormalised ground state energy.

\section{Cosmic string with finite thickness }\label{Sec4}
We use the metric given by Eq. \Ref{m1_1}, \Ref{m2_1}. Again, the
coordinate $z$ will be dropped because we work in (2+1)
dimensions. The curvature scalar is
\be\label{cusc}
{\rm R}= -\frac{2\epsilon^2}{\rho_0^2}\ ,
\ee
and we allow for $\xi\ne\frac16$. 
By means of the ansatz $(n = 0, \pm 1, \pm 2 , \dots)$
\be\label{ansatz2}
\Phi = e^{in\varphi }g^{-1/4}{\cal R} = e^{in\varphi } 
   \left\{ 
          \begin{array}{l}
          {\cal R}_{in}(\rho)/\sqrt{\frac{\rho_0}{\epsilon} 
           \sin (\frac{\epsilon \rho}{\rho_0})}\ ,\ 
           \rho \in [0,\rho_0] \\
           {\cal R}_{out}(r)/\sqrt{\frac{r}{\nu}} \ ,\ r 
           \in [r_0, \infty ) \ ,
          \end{array}
   \right. 
\ee
we arrive at the equations
\bea
\left\{\frac{d^2}{d\rho^2} - \frac{\epsilon^2 [n^2 -1/4]}{
\rho_0^2\sin^2 (\epsilon\rho /\rho_0)} + \frac{\epsilon^2}{4\rho_0^2
}(1 - 8\xi) + \lambda^2 \right\}{\cal R}_{in} &=& 0\ ,\ 
\rho \in [0,\rho_0] \label{in} \\  
\left\{\frac{d^2}{dr^2} - \frac{n^2 \nu^2 - 1/4}{r^2} + \lambda^2
\right\}{\cal R}_{out}&=& 0 \ ,\ r \in [r_0, \infty )  \label{out} 
\eea
for the radial functions.  Both this equations may be solved
exactly. Indeed, the solution of the radial equation outside of the
string (\ref{out}) can be written as
\be {\cal R}_{out} = \frac{i}{2}\left[ f_n(\lambda )
H^-_{n\nu}(\lambda r) - f_n^* (\lambda) H^+_{n\nu}(\lambda r)\right]\
, \ee
 where $f_n$ is the Jost function and 
\be 
H^{\pm}_{n\nu}(\lambda r) = \pm i \sqrt{\frac{\pi\lambda
    r}{2}} H^{{(1)},{(2)}}_{n\nu}(\lambda r)\ 
\ee 
are combinations of the Hankel functions. The solution regular at the origin 
of the radial equation (\ref{in}) inside the string is
\[
 {\cal R}_{in} =
e^{i\frac{\pi}{2}n(\nu -1)} \sqrt{\frac{\pi}{2}\sin\frac{ \epsilon\rho}{
    \rho_0}} \left(\frac{ \lambda\rho_0}{\epsilon}\right)^{n+1/2}{\rm
  P}^{-n}_\alpha \left[\cos\frac{\epsilon\rho}{\rho_0}\right]\ 
\]
with 
\[
 \alpha = -\frac{1}{2} + \frac{1}{2}\sqrt{1 +
  \frac{4\lambda^2\rho_0^2}{ \epsilon^2} - 8\xi}\   
\]
and ${\rm P}^n_\alpha$
is the Legendre function. These solutions must obey the 
  matching conditions
\[
 {\cal
  R}_{in}(\rho_0) = {\cal R}_{out}(r_0)\ ,\ {\cal R}'_{in}(\rho_0) = {\cal
  R}'_{out}(r_0) 
\]
on the surface of the string. 
From this we get the following formula for the Jost function
\beao f_n(\lambda )&=&-e^{i\frac{\pi}{2}n(\nu -1)}\frac{i\pi}{2}
  \frac{\sin
  \epsilon}{\sqrt{\cos\epsilon}}\left(\frac{\lambda\rho_0}{\epsilon}\right
  )^{n+1} \\ &&\times \left\{{H^{(1)}_{n\nu}}'(\lambda r_0){\rm
  P}^{-n}_\alpha [\cos\epsilon] + H^{(1)}_{n\nu}(\lambda r_0){{\rm
  P}^{-n}_\alpha}' [\cos\epsilon]\frac{\epsilon \sin\epsilon
  }{\lambda\rho_0}\right\} \ .  \eeao
Here the primes denote derivatives with respect to the
arguments. Taken on the imaginary axis this Jost function reads
 \bea f_n(ik)&=&-\frac{\sin\epsilon}{\sqrt{\cos\epsilon}}
\left(\frac{k\rho_0}{\epsilon}\right)^{n+1} \label{ImJost} \\
&&\times \left\{K'_{n\nu}(kr_0){\rm P}^{-n}_\alpha [\cos\epsilon] +
K_{n\nu}(kr_0){{\rm P}^{-n}_\alpha}' [\cos\epsilon]\frac{\epsilon
\sin\epsilon }{k\rho_0}\right\} \ . \nonumber \eea
Note that this function doesn't have zeros for $k\in[0,\infty)$, i.e.,
 there are no bound states. This can be checked by inspection of eq.
 \Ref{ImJost}. The Jost function has the following asymptotics for
 large resp.  small $k$:
\bea f_n(ik)\strut_{k \to \infty} &\sim
&\exp\{-k\rho_0(\frac{r_0}{\rho_0} - 1)\} \ , \nonumber \\
f_n(ik)\strut_{k \to 0} &\sim & k^{-n(\nu -1)}\ ,\ f_0(ik)\strut_{k
\to 0} \sim -\ln k\ .
\label{limit}
\eea
Using the formula 
\be
\lim_{t\to\infty} t^n {\rm P}^{-n}_t [\cos \frac{x}{t}] = J_n(x) \ ,
\ee
the Minkowskian limit ($\nu \to 1$) for the Jost function 
\be
\lim_{\nu\to 1} f_n(\lambda ) = 1\ 
\label{Minkowskilimit}
\ee
can be checked.

Proceeding as it was done in the case of the infinitely thin string we
obtain the following expression for the ground state energy in  $(2+1)$
dimensions 
\beao
E_0 &=& -M^{2s}\frac{\cos\pi s}{2\pi} \sum_{n
  = 0}^{+\infty } d_n \int_m^\infty dk [k^2 - m^2]^{1/2 - s}\\
&&\times\frac{\partial }{\partial k} \ln \left\{k^{-n}\left[f_n(ik)
H^-_{n\nu}(ikR) - f_n^* (ik) H^+_{n\nu}(ikR) \right]\right\} \ 
\eeao
and,  using 
\[
K_{\mu}(z\e^{-i\pi})=\e^{-i\mu z}K_{\mu}(z)-i\pi I_{\mu}(z)
\]
after obvious rearrangements we get
\beao
E_0&=&  
-M^{2s}\frac{\cos\pi s}{2\pi} \sum_{n = 0}^{+\infty } d_n
\int_m^\infty dk [k^2 - m^2]^{1/2 - s}\\&& \times\frac{\partial }{\partial k}
\ln\left\{k^{-n}\left[f_n(ik) I_{n\nu}(kR) - \widetilde{f}_n (ik)
K_{n\nu}(kR) \right]\right\} \\ 
    &=& 
-M^{2s}\frac{\cos\pi s}{2\pi} \sum_{n = 0}^{+\infty } d_n
\int_m^\infty dk [k^2 - m^2]^{1/2 - s} \frac{\partial }{\partial
k}\ln k^{-n}f_n(ik) I_{n\nu}(kR) \nonumber \\ 
    &&-
M^{2s}\frac{\cos\pi s}{2\pi} \sum_{n = 0}^{+\infty } d_n
\int_m^\infty dk [k^2 - m^2]^{1/2 - s} \frac{\partial }{\partial k} 
\ln\left[ 1 - \frac{\widetilde{f}_n (ik)}{f_n(ik)}
\frac{K_{n\nu}(kR)}{I_{n\nu}(kR)} \right] \nonumber .
\eeao

Now the contribution of the last line is exponentially small for
$R\to\infty$. The contribution of the preceeding line can be written as the
sum
\be\label{sum}
\E_{0}=\E_{0}^{\rm thin}+\E_{0}^{\rm int}
\ee
of the contribution $\E_{0}^{\rm thin}$ \Ref{zeta2} of the infinitely thin
string considered in the preceeding section and the contribution of the
interior structure of the string
\be\label{Eint}
E^{\rm int}_0 = - M^{2s}\frac{\cos\pi s}{2\pi} \sum_{n
= 0}^{+\infty } d_n \int_m^\infty dk [k^2 - m^2]^{1/2 - s}
\frac{\partial }{\partial k} \ln k^{n(\nu -1)}f_n(ik) \ .
\ee 
After the work done in section 3 it is just this contribution which remains
to be calculated.

\section{Approximation of small angle deficit}\label{Sec5}

To calculate $E^{\rm int}_0$ \Ref{Eint} and the corresponding
renormalised ground state energy the analytic continuation in $s$ to
$s=0$ must be performed. To this end the knowledge of the uniform
asymptotic expansion of the Jost function $f_{n}(ik)$ \Ref{ImJost} for
$n\to\infty$, $k\to\infty$, $n/k$ fixed is required. Now, althought
this Jost function is known explicitely in terms of Bessel functions
and the Legendre function, this task is not easy to perform.  The
point is that the asymptotic for $\alpha\to \infty$ and $n\to\infty$
of the Legendre function $P_{\alpha}^{-n}$ is quite complicated to
handle.  Therefore we restrict ourselfs to the easier tractable case
of a small angle deficit resp. mass density of the string, i.e., to
the case $\ep \ll 1$.  Than in the radial equation for the interior of
the string
\be\label{RadialEquation}
\left\{\frac{d^2}{d\rho^2} - \frac{n^2 - 1/4}{\rho^2} + \lambda^2
 - U(\rho )\right\}{\cal R}_{in}= 0\ ,\ \rho \in [0,\rho_0]\ ,
 \ee
with the potential (using \Ref{cusc})
\[
U(\rho)={1\over \rho^{2}}\left(\left(n^{2}-\frac14\right)
  \left({\theta^{2}\over\sin^{2}\theta}-1\right)+
  \theta^{2}\left(2\xi-\frac14\right)\right)
\]
($\theta=\ep\rho/\rho_{0}$) we approximate
\be\label{apu}
U(\rho)=U_{0}+O(\ep^{4})
\ee
with
\[
U_{0}= {e^{2}\over 3\rho_{0}^{2}}(n^{2}+6\xi-1).
\]
By this,  Eq. \Ref{RadialEquation} can be solved in terms of Bessel functions 
\[
{\cal R}_{in} = e^{i\frac{\pi n}{2}(\nu -
1)}\left(\frac{\lambda}{\lambda_n}\right)^{n+1/2}\sqrt{ 
\frac{\pi}{2} \lambda_n \rho} \ J_n(\lambda_n \rho) \ ,\ 
\]
($\lambda_n = \sqrt{\lambda^2 - U_0}$) and the corresponding Jost function
reads for small $\ep$
\beao
f^{\rm se}_n(\lambda ) &=& i\frac{\pi}{2}e^{i\frac{\pi n}{2}(\nu - 1)}
\left( \frac{\lambda }{\lambda_n }\right)^n\left\{
\frac{1}{2} \left[\sqrt{\frac{r_0}{\rho_0}} -
\sqrt{\frac{\rho_0}{r_0}}\right] J_n(\lambda_n \rho_0)
H_{n\nu}^{(1)}(\lambda r_0)\right. \\ 
&&+ \left. \sqrt{r_0\rho_0} \left[ \lambda_n
J_n'(\lambda_n \rho_0) H_{n\nu}^{(1)}(\lambda r_0) - \lambda
J_n(\lambda_n \rho_0) {H_{n\nu}^{(1)}}'(\lambda
r_0)\right]\right\}\ . \nonumber
\eeao
On the imaginary axis we get
\bea\label{fnik}
f^{\rm se}_n(ik ) &=& \left( \frac{k }{k_n }\right)^n\left\{
\frac{1}{2} \left[\sqrt{\frac{r_0}{\rho_0}} -
\sqrt{\frac{\rho_0}{r_0}}\right] I_n(k_n \rho_0) K_{n\nu}(k
r_0)\right.  \\ 
&&+ \left. \sqrt{\frac{r_0}{\rho_0}} \left[ k_n\rho_0 I_n'(k_n
\rho_0) K_{n\nu}(k r_0) - k\rho_0 I_n(k_n \rho_0) 
K_{n\nu}'(k r_0)\right]\right\}\ ,   \nonumber
\eea
with $r_0/\rho_0 = \tan\epsilon /\epsilon $ and
$k_n = \sqrt{k^2 + \frac{\epsilon^2}{3\rho_0^2}(n^2 + 6\xi - 1)}$. It is easy
to verify that it obeys the limits \Ref{limit} and \Ref{Minkowskilimit}. 

Note that the approximate potential $U$ \Ref{apu} is constant and that
$U_{0}$ may take negative values. Therefore bound states could
occur. However, due to the conical structure of the exteriour part of
the space (see $\nu$ in the index of the Hankel functions) and the
corresponding relations between the parameters, in fact no bound
states occur. Correspondingly, it can be shown that the Jost function
\Ref{fnik} doesnt have zeros for $k\in[0,\infty)$.
 
Now we insert this approximate Jost function into $\E_{0}^{\rm int}$
\Ref{Eint} and have to perform the analytic continuation in $s$ to
$s=0$. For this reason we use the uniform asymptotic expansion of
\Ref{fnik} which can now be obtained by simply inserting the
corresponding expansion \Ref{ufe} of the Bessel functions. We define
\be\label{UEJ}
f_n^{\rm se,~as}(ik) = \frac{z^{n\nu}}{y^n} e^{-n \nu \eta
(z_\nu ) + n \eta (y)} 
 \left( 1 + \epsilon^2 \frac{t}{24n} \right)
\ee
with
\beao
z&=&\frac{k\rho_0}{n}\ ,\ z_\nu=\frac{kr_0}{n\nu}\ ,\ t =
(1+z^2)^{-1/2}\ ,\\  
\eta (z)&=&\sqrt{1+z^2} + \ln\frac{z}{1+\sqrt{1+z^2}} \ ,\ y =
\sqrt{z^2 + \frac{\epsilon^2 }{3} - \frac{\epsilon^2 }{n^2 }\frac{1 -
6\xi}{3}}\ .  
\eeao
In general, the expansion \Ref{UEJ} must include all terms up to
$n^{-2}$. But that contribution doesn't appear (together with
$n^{-4}$, $n^{-6}$, ...).  Next we divide the expression for
$\E_{0}^{\rm int}$ into two parts
\be\label{div}
E^{\rm int}_0 = E^{\rm int}_{0,\rm as} + E^{\rm int}_{0,\rm fin}\ . 
\ee
simply subtracting and adding $\ln k^{n(\nu -1)}f_n^{\rm se,~as}$ in
\Ref{Eint}. Here
\be\label{Eintas}
E^{\rm int}_{0,\rm as} = - M^{2s}\frac{\cos\pi s}{2\pi} \sum_{n
= 0}^{+\infty } d_n \int_m^\infty dk [k^2 - m^2]^{1/2 - s}
\frac{\partial }{\partial k} \ln k^{n(\nu -1)}f_n^{as}(ik) \ ,
\ee 
is called the 'asymptotic' part which still requires the analytic
continuation to be done and
\be\label{Eintfin}
E^{\rm int}_{0,\rm fin} = \frac{-1}{2\pi} \sum_{n
= 0}^{+\infty } d_n\!\!\int_m^\infty\!\! dk \sqrt{k^2 - m^2}
\frac{\partial }{\partial k} \left[ \ln k^{n(\nu -1)}f_n(ik) -
\ln k^{n(\nu -1)}f_n^{as}(ik)\right] \ 
\ee 
is called the 'finite' part. In it the analytic continuation could be
performed under the sign of the integral and the sum because they are
convergent.

In $E^{\rm int}_{0,\rm as}$ \Ref{Eintas} the integrand can be expanded in
powers of $\ep$. Than the integration over $k$ and the summation over $n$ can
be carried out explicitely using the same method as in section 3. We
obtain 
\bea \label{ZetaIntN}
E^{\rm int}_{0,\rm as}|_{s\to 0} &=& \frac{\epsilon^2}{2\pi \rho_0} 
           \left\{ \frac{\pi}{24}\beta^3 - \left[\frac{\pi }{12} - 
\frac{1 - 6\xi}{12}\pi \right] \beta -\left[ \frac{1}{720} +
\frac{1-6\xi}{72} \right]\frac{\pi}{\beta}  
           \right. \nonumber \\ 
&&+        \left[ 
           \frac{1}{6}\beta^2 - \frac{1}{24} + \frac{1 - 6\xi}{6} 
           \right] \ln (1 - e^{-2\pi \beta })
           \\
&&-        \left. \frac{\pi}{3}\beta^2
           \left[ 3 \widetilde{Q}_1(\beta ) - 3\widetilde{Q}_2(\beta ) 
            + \widetilde{Q}_3(\beta )\right] - 
                 \frac{1 - 6\xi }{3} \pi \widetilde{Q}_1(\beta )
               \right\}\ , \nonumber 
\eea
with the notation $\beta=m\rho_{0}$ and $\tilde{Q}_{a}$ defined by
\Ref{Qtilde}.  We remark that $\tilde{Q}_{a}(\beta)$ are exponentially
decreasing for $\beta\to\infty$.

At this place we can determine the heat kernel coefficients for the
thick string. We have to consider the asymptotic expansion of
$\E_{0}^{\rm int}$ for large $m$. The nondecreasing contributions may
be contained only in $E_{0,\rm as}^{\rm int}$ \Ref{Eintas} and can be
read off from Eqs. \Ref{ZetaIntN} and \Ref{ehk}. They are
\be\label{bsk}
B_{0}^{\rm int}=-\ep^{2}{\pi\rho_0^{2}\over4}, ~~~B_{1}^{\rm
  int}=-2\pi\xi\ep^{2}\, .
\ee
Note that the coefficient $B_{1/2}$ is zero. This was to be expected
because the background is smooth enougth not to allow for boundary
dependent coeficients in this order. Also, the coefficient $B_{3/2}$
is zero. This is in the given order in $\ep$ and follows simply from
dimensional reasons. In higher orders in $\ep$ it may be nonzero like
further coefficients of higher half integer order.

Now, by means of \Ref{sum} we have to take the contributions to the
heat kernel coefficients of the infinitely thin string \Ref{hkks} and
that of the 'interior', Eq.  \Ref{bsk} together. These coefficients
can be compared with that following from the general formulas. For
instance, from
\bea\label{B0th}
B_{0}=\int\limits_{V}\d
V&=&2\pi\int\limits_{0}^{\rho_{0}}{\rho_{0}\over\ep}\sin{\ep\rho\over\rho_{0}}
d\rho +{2\pi\over\nu}\int\limits_{r_0}^{R}r\d r  \nn \\
&=&{\pi R^{2}\over\nu}-{\ep^{2}\pi\rho_{0}^{2}\over 4}+O(\ep^{4})\ ,
\eea
we obtain the boundary dependent contribution (cf. Eq. \Ref{hkks}) and
$B_{0}^{\rm int}$ \Ref{bsk}.  For the coefficient $B_{1}$ we have
\be\label{B1th}
B_{1}={1-6\xi\over 6}\int\limits_{V}{\rm R}\d V={1-6\xi\over
  3}{ \ep^{2}}\pi+O(\ep^{4}).
\ee
Now, in the given approximation it holds
\[
{\pi\over 3}\left(\nu-{1\over\nu}\right)={\pi\over 3}\ep^{2}+O(\ep^{4}).
\]
Therefore, from $B_{1}$ \Ref{hkks} resp. \Ref{B-1} of the thin string and
$B_{1}^{\rm int}$ \Ref{bsk} we get
\be\label{}
B_{1}=\frac\pi 3 \left(\nu+{1\over\nu}\right)-2\pi\xi\ep^{2}={2\pi\over 3\nu}
+ \ep^{2}\pi \frac{1-6\xi}{3}+ O(\ep^{4}).
\ee
We note that the Kac term disappeared. The first term in the last line is due
to the boundary at $r=R$ and the second is the genuine contribution of the
thick string. It vanishes in case of a conformal coupling. 
 
To proceed, we define $E_{0}^{\rm ren}$ by means of \Ref{eren} using
the coefficients $B_{0}$ \Ref{B0th} and $B_{1}$ \Ref{B1th} in the
definition of $E_{0}^{\rm div}$ \Ref{ehk}. Note that the coefficients
with half integer numbers are pure boundary dependent contributions
resulting from the boundary conditions at $r=R$.

In fact, the renormalization in $E_{0}^{\rm int}$ is reduced to
dropping the nondecreasing for $m\to\infty$ contributions in $E_{0,\rm
as}^{\rm int}$. Therefore we obtain for the complete renormalized
ground state energy
\bea\label{interior} 
E^{\rm ren}_0 &=& \frac{\epsilon^2}{2\pi \rho_0} 
           \Bigg\{   
               \left[ 
           \frac{1}{6}\beta^2 - \frac{1}{24} + \frac{1 - 6\xi}{6} 
           \right] \ln (1 - e^{-2\pi \beta })
          \nn \\
&&-          \frac{\pi}{3}\beta^2
           \left[ 3 \widetilde{Q}_1(\beta ) - 3\widetilde{Q}_2(\beta ) 
            + \widetilde{Q}_3(\beta )\right] - 
                 \frac{1 - 6\xi }{3} \pi \widetilde{Q}_1(\beta ) 
          \\ 
&&      - \sum_{n=0}^{\infty }d_n
           \int_\beta^\infty dx \sqrt{x^2 - \beta^2 } 
\frac{\partial }{\partial x} F_n -\left[ \frac{1}{720} + \frac{1 -
6\xi}{6}\right]\frac{\pi}{\beta} 
           \Bigg\}\ . \nonumber 
\eea
Here, the notation
\[
F_{n}={\ln f_n(ix ) - \ln f_n^{as}(ix) \over \epsilon^2}
\]
is introduced. 

Some further work is necesary. In $F_{n}$ the Jost functions have
 still to be expanded for small $\ep$. For $f_{n}(ik)$ this can be
 done using the formula \cite{Abramowitz}
\[
\frac{\partial K_p(x)}{\partial p}|_{p=n} = \frac{n!}{2} \left(
\frac{x}{2} \right)^{-n} \sum_{l=0}^{n-1}\left( \frac{x}{2}
\right)^{n} \frac{K_l(x)}{(n-l)l!}\ ,
\]
for $f_{n}^{\rm as}(ik)$ this expansion is a simple task. Finally
we obtain 
\beao {F_n}_{|_{{\ep=0}}} &=& \left( \frac{(n + 1)^2}{6} - \frac{x^2}{3}
\right) I_n K_n + \left( \frac{n^2}{6} - \frac{x^2}{3} \right) I_{n+1}
K_{n-1}+ \frac{x}{3} I_{n+1} K_n
\nonumber \\
&&+ \frac{nn!}{4} \sum_{l=0}^{n-1}\left( \frac{x}{2}\right)^{l-n}
\frac{x(I_{n+1} K_l + I_n K_{l+1})}{(n-l)l!} + \frac{n}{2} \left( K_0
  n!I_n\left( \frac{2}{x} \right)^n + \ln x \right)
\nonumber \\
&&+ \left.\frac{nn!}{2} \sum_{l=1}^{n-1}\left( \frac{x}{2}\right)^{l-n}
  \frac{I_n K_l}{l!}\right|_{n\geq 2} + \frac{n}{3} \sqrt{1 + \frac{x^2}{n^2}}
-
\frac{n}{6} \frac{1}{1 + \sqrt{1 + \frac{x^2}{n^2}}}\nn \\
&& - \frac{n}{2} \ln \left( 1 + \sqrt{1 + \frac{x^2}{n^2}} \right)
- \frac{1}{24n} \frac{1}{\sqrt{1 + \frac{x^2}{n^2}}} \nn \\
&&- \frac{1 - 6\xi }{6} \left[ I_n K_n + I_{n+1} K_{n-1} -
  \frac{1}{n}\frac{1}{1 + \sqrt{1 + \frac{x^2}{n^2}} } \right]\ .  
\eeao 
This expression has to be used in $E^{\rm ren}_0$ \Ref{interior}.

In writing
\be\label{Fn}
E^{\rm ren}_{0}={\ep^{2}\over 2\pi\rho_{0}}G(\beta)+O(\ep^{4})
\ee
it is in fact $G(\beta)$, a function of one variable, which must be
calculated. We did this task numerically. After a carefully
examination we came to the result
\be\label{Gistnull}
G(\beta)=0 \ ,
\ee
for arbitrary $\xi$; the details are given in the appendix.

\section{Conclusions}\label{Sec6}

In the preceeding sections we worked out methods suited for the
calculation of the ground state energy of a massive scalar field in
the background of a cosmic string. The main emphasis was on a string
of finite thickness. We used the standard renormalization scheme,
i.e., we calculated and subtracted the contributions of the first few
heat kernel coefficients. Thereby the normalisation condition, stating
that the renormalised ground state energy must vanish for a large mass
of the quantum field, is imposed.

As a part of these calculations we first considered the infinitely
thin string in detail. Using explicit formulas we reobtained the known
heat kernel coefficients. These are the coefficients due to the
boundary of a large cylinder at $r=R$, which was introduced in order
to render the eigenvalues discrete, and the topological Kac term. In
the sense of the renormalisation used we got the result that the
renormalised ground state energy of the pure string, i.e., when
removing the outer boundary ($R\to\infty$), is zero.

Then the same problem is formulated for a string with finite
thickness. However, the complete calculation suffers still from
mathematical difficulties. Therefore the approximation of a small mass
density $\ep$ of the string was introduced. Having in mind the
smallness of $\ep\sim 10^{-3}$ in cosmological applications, this is
at once physically motivated.

We note that in this approximation an alternative calculation should
be possible, namely the use of a perturbation theory in the mass
density as it was done for the calculation of the Casimir force
between two cosmic strings in \cite{bordag90}.

In this approximation of a small mass density, in order $\ep^{2}$,
first the heat kernel coefficients were calculated. They are checked
to coincide with that following from general formulas. We remark that
the Kac term disappeares and that for a conformal coupling ($\xi=1/6$)
there are no counterterms required besides that which follow from the
boundary at $r=R$. From this it is clear that the Kac term is due to
the singular behaviour of the metric of the thin string at the
origin. This can be understood from another point of view
too. Consider the vacuum expectation value of the energy density in
the background of the thin string. For dimensional reasons it behaves
like $r^{-2}$ near the string and, therefore, cannot be integrated
over $r$ near the origin. Now, if introducing a suitable
regularisation, zeta functional regularisation for instance, it
becomes possible to integrate over $r$. As a result, when removing the
regularisation, an additional divergence occurs which is just the Kac
term.

After performing the renormalisation we calculated numerically the
ground state energy in the background of the thick cosmic string in
order $\ep^{2}$ of a small mass density. The result is zero with the
reasonable precision of $10^{-7}$ independently of the parameter
$\xi$. Thereby a nontrivial compensation between different
contributions occured.

This result that the vacuum of a scalar field is not disturbed by a
cosmic string is quite remarkable and seems not to have analogues in
other configurations.  For instance, we do not see any symmetry or
invariance arguments for this result although they should be there.

Perhaps, there is some relation to the result of Brevik and Jenesen
\cite{brevikjensen} indicating the absence of particle production in the
formation of a cosmic string.

Further work is necessary. For instance, the result should be extended
to the (3+1) dimensional case, to higher spin fields and to mass
densities which are not small.

\section*{Acknowledgement}
The authors would like to thank K. Kirsten and D. Fursaev for discussions. \\
NK is supported in part by Russian Found of Fundamental
Research (grant No 97-02-16318). 
\section*{Appendix} 
Here we consider the analytical and numerical analysis of the function
$G(\beta)$ defined in Eq. \Ref{Gistnull}
\bea  \label{bg1}
G(\beta )
&=& \left[ 
          \frac{1}{6}\beta^2 - \frac{1}{24} + \frac{1 - 6\xi}{6} 
    \right] 
    \ln (1 - e^{-2\pi \beta }) - \frac{\pi}{3}\beta^2
    \left[ 3 \widetilde{Q}_1(\beta ) - 3\widetilde{Q}_2(\beta ) 
           + \widetilde{Q}_3(\beta )
    \right]\nn \\
&&-    \frac{1 - 6\xi }{3} \pi \widetilde{Q}_1(\beta ) - 
       \sum_{n=0}^{\infty }d_n
       \int_\beta^\infty\! dx \sqrt{x^2 - \beta^2 } 
       \frac{\partial F_n}{\partial x} - 
       \left[ \frac{1}{720} + \frac{1 - 6\xi}{72}
       \right]
\frac{\pi}{\beta}.
\eea
Obviously we have $G(\beta )_{\beta \to \infty} \to 0$ and the domain of
interest is the neighborhood of zero: $\beta \sim 0$. For numerical
simulations the above formula is more suitable for $\beta \geq 1$. In the
opposite case, $\beta < 1$, there exist at first sight a logarithmic
singularity for small $\beta$ : $\ln (1 - \exp (-2\pi \beta )) \sim \ln 2\pi
\beta$. But this singularity is canceled with that in the contribution of
$(n=0)$ in the series in \Ref{bg1}. For numerical calculations it is more
suitable to cancel this singularity in manifest form. For this reason we
divide this term into two parts, 
\be
 \int_\beta^\infty dx \sqrt{x^2 - \beta^2
  } \frac{\partial }{\partial x} F_0 = \int_\beta^1 dx \sqrt{x^2 - \beta^2 }
\frac{\partial }{\partial x} F_0 + \int_1^\infty dx \sqrt{x^2 - \beta^2 }
\frac{\partial }{\partial x} F_0\ , 
\ee
where 
\beao
F_0   &=& \Phi_0 - \Phi_0^{as}, \\
\Phi_0 &=& \frac{1 - x^2}{6} I_0K_0 - \frac{x^2}{3}I_1K_1 - \frac{x^2}{6}
I_0K_2 - \frac{1 - 6\xi}{6} (I_0K_0 + I_1K_1),\\
\Phi_0^{as} &=& -\frac{x}{3} + \frac{1}{24x} - \frac{1 - 6\xi}{6x}\ .  
\eeao
The integral %
\be 
\int_\beta^1 dx \sqrt{x^2 - \beta^2 } \frac{\partial
  }{\partial x} \Phi_0^{as}\  
\ee
 may be found in manifest form. Thereby we
arrive at the following expression for the case $\beta <1$ 
\bea G(\beta ) &=&
\left[ \frac{1}{6}\beta^2 - \frac{1}{24} + \frac{1 - 6\xi}{6} \right] \ln
\left[\frac{1 - e^{-2\pi \beta }}{\beta}(1 + \sqrt{1 - \beta^2)}\right]
                   \label{bk1} \\
&&- \frac{\pi}{3}\beta^2 
                         \left[ 
                               3 Q_1(\beta ) - 
                               3 Q_2(\beta ) 
                               + Q_3(\beta )
                        \right] - 
                        \frac{1 - 6\xi }{3}\pi Q_1(\beta ) \nonumber \\  
           &&+ \frac{\pi\beta}{24} + \zeta'_R(-2) - 
               \left[\frac{1}{8} + \frac{1 - 6\xi}{6}\right]
               \sqrt{1 - \beta^2} - \int_\beta^1 dx \sqrt{x^2 - \beta^2 } 
           \frac{\partial }{\partial x} \Phi_0 \nonumber \\
           &&- \int_1^\infty dx \sqrt{x^2 - \beta^2 } 
           \frac{\partial }{\partial x} F_0 - 
           \sum_{n=1}^{\infty }2
           \int_\beta^\infty dx \sqrt{x^2 - \beta^2 } 
           \frac{\partial }{\partial x} F_n  
           \ .  \nonumber
\eea
with the notation
\be
Q_{n}(\beta ) = \frac{1}{\beta^{n}} \int_0^\beta \frac{dt
t^{n}}{e^{2\pi t} - 1}\ .
\ee

It is necessary to stress that formula (\ref{bk1}) is only a different
representation of $G(\beta)$ \Ref{bg1} which we made in order to avoid
the logarithmic contribution in individual terms.  Next we observe
that the series over $n$ is slowly convergent and quite a large number
of terms must be taken into account

Now let us consider some analytical properties of $G(\beta)$. This function
doesn't have a linear term in the expansion for small $\beta$, 
\bd 
G(\beta )
= G(0) + O(\beta^2)\ .  
\ed
In order to argue this statement let's consider
the part in (\ref{bk1}) which contains the integral 
\bea D(\beta ) &=&
\int_\beta^1 dx \sqrt{x^2 - \beta^2} \frac{\partial \Phi_0 }{\partial x} +
\int_1^\infty dx \sqrt{x^2 - \beta^2} \frac{\partial F_0 }{\partial
x} \nonumber \\ 
&+&
\sum_{n=1}^{\infty }2 \int_\beta^\infty dx \sqrt{x^2 - \beta^2} \frac{\partial
  F_n }{\partial x}\ . \label{D} 
\eea
The function $\Phi_0$ has the following expansion for small $x$ (using the
power series expansion of the Bessel functions)
\bd \Phi_0 = \ln \frac{x}{2} \sum_{k=0}^{\infty} C_k x^{2k} +
\sum_{k=0}^{\infty} \widetilde{C}_k x^{2k} \ .  
\ed
It is only the zeroth term in the first sum, $C_0\ln x/2$, which
deliveres a linear contribution to the first integral in
(\ref{D}). All other terms contribute higher powers in $\beta$. Thus
\bd \int_\beta^1 dx \sqrt{x^2 - \beta^2} \frac{\partial \Phi_0
}{\partial x} = \int_0^1 dx x \frac{\partial \Phi_0 }{\partial x} -
\frac{\pi}{2} C_0 \beta + O(\beta^2)\ \ed
with $C_0 = (1 - 6\xi)/6 - 1/6 = -\xi$.  

In the second integral in (\ref{D}) we can expand the integrand for
$\beta^2/x^2 \ll 1$ obtain
\bd
\int_1^\infty dx \sqrt{x^2 - \beta^2} \frac{\partial
F_0 }{\partial x} = \int_1^\infty dx x \frac{\partial
F_0 }{\partial x} + O(\beta^2)\ . 
\ed
In the last term in (\ref{D}) we use the uniform expansion of $F_n$
(\ref{UEJ}) which starts from the third power of $1/n$. All integrals
can be calculated in closed form, their expansion starts from the
second order in $\beta$. Therefore
\bd
\sum_{n=1}^{\infty }2 \int_\beta^\infty dx 
\sqrt{x^2 - \beta^2} \frac{\partial F_n }{\partial x} =
\sum_{n=1}^{\infty }2 \int_0^\infty dx x\frac{\partial F_n }{\partial
x} + O(\beta^2)\ . 
\ed 
Puting together all three parts we obtain   
\be
D(\beta ) = D(0) +\frac{\pi}{2}\xi\beta + O(\beta^2)\ . \label{DExp}
\ee
The expansion of the remaining terms in (\ref{bk1}) contains the same linear
term
\bd
const + \frac{\pi}{2}\xi\beta + O(\beta^2)\ ,
\ed
which is canceled by that in (\ref{DExp}). Therefore the expansion
of $G(\beta )$ starts at least from the second power of $\beta$. 
\begin{figure}[tbh]
  \centerline{
  \psfig{figure=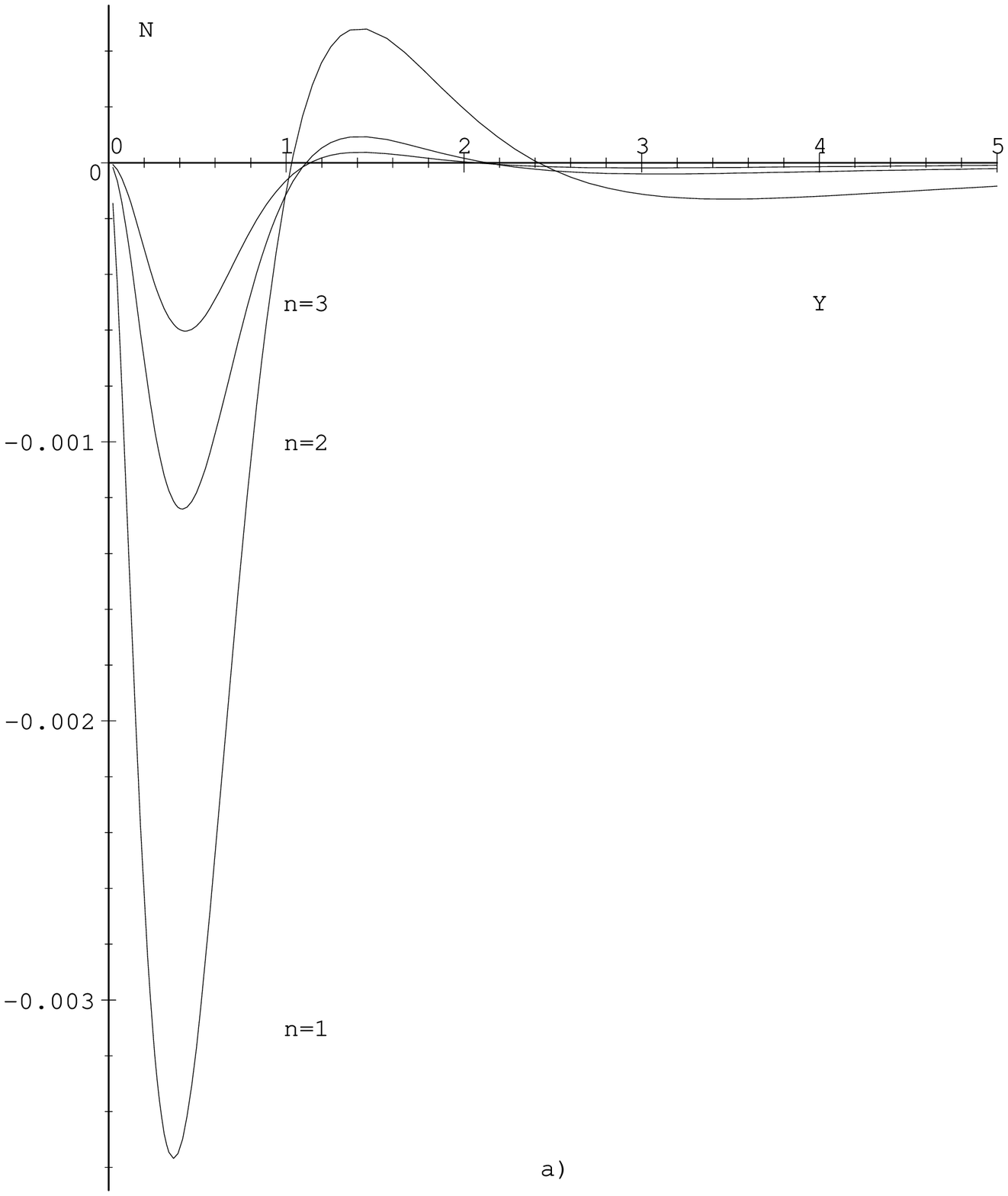,angle=0,height=12cm}
\psfig{figure=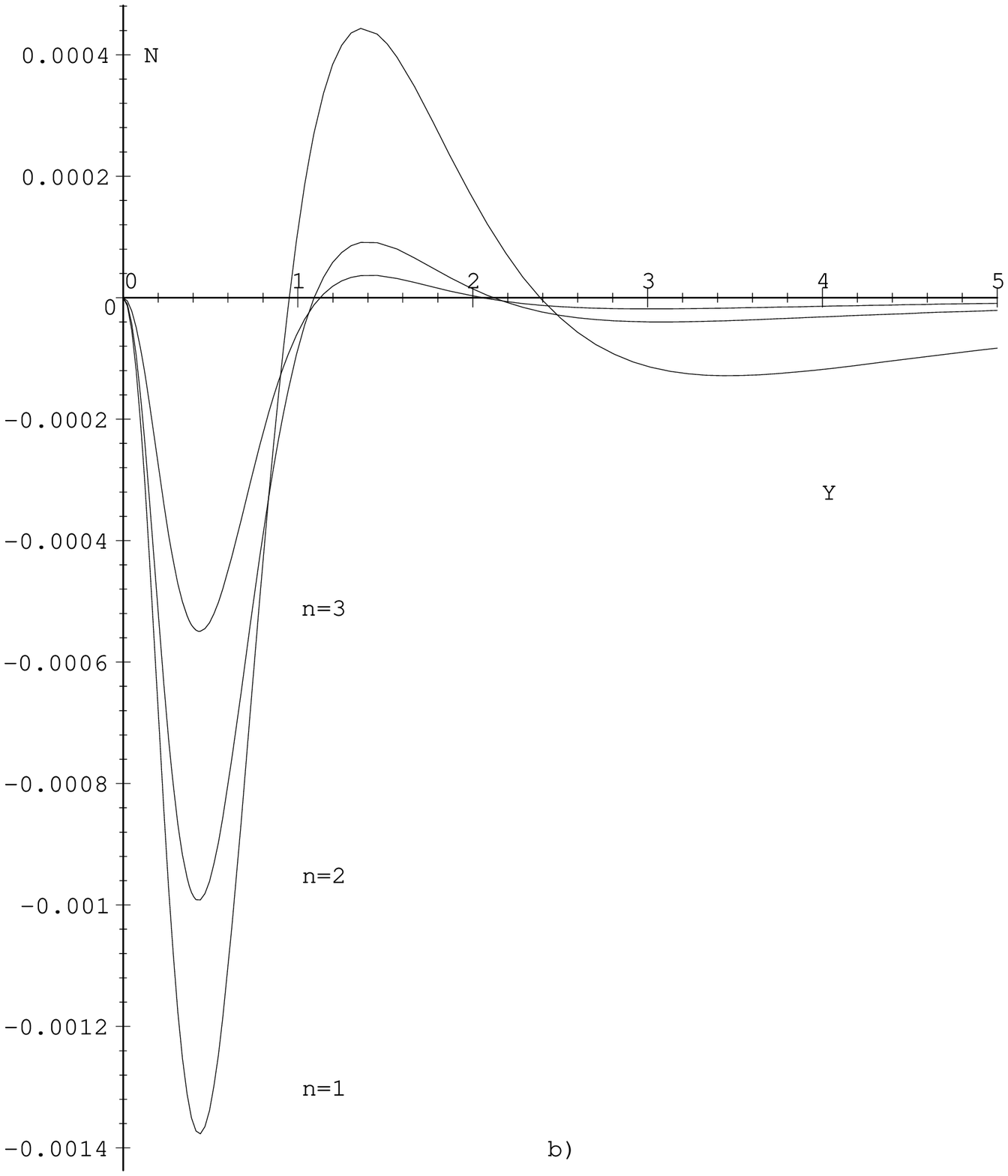,angle=0,height=12cm}}
\caption{Plots of the integrand $N_n = 2ny \frac{\partial }{\partial
y} F_n ((n^2y^2 + \beta^2)^{1/2})$ for a) $\beta =0$ and b) $\beta
=0.4$ and $n=1,2,3$}\label{N}
\end{figure}

For the numerical analysis of the series in $G(\beta )$ first of all we
replace the integration variable $x\to y = \sqrt{x^2 - \beta^2}/n$.
Then we obtain
\bd
\sum_{n=1}^{\infty } \int_\beta^\infty  2\sqrt{x^2 - \beta^2 }
\frac{\partial }{\partial x} F_n(x) dx= 
\sum_{n=1}^{\infty } \int_0^\infty 2ny \frac{\partial }{\partial y} 
F_n(\sqrt{n^2y^2 + \beta^2})dy\ . 
\ed
Some first integrands $N_n = 2ny \frac{\partial }{\partial y} 
F_n(\sqrt{n^2y^2 + \beta^2})$ are plotted in Fig.\ref{N} for $\beta=0$ and
for $\beta = 0.4$. As it is seen from the figures, all functions
$N_n(y)$ have quite large variations near the origin and decrease as
$1/y^3$ for large $y$. 

For higher $n$, ($n>3$), the Bessel functions entering $F_{n}$, have been
substituted by their unifom asymptotic expansions   whereby the first 13
terms were taken. The error caused by this approximation is smaller than
$10^{-7}$. Than the integral and the sums can be carried out explicitely. For
this task Maple was used. 

With this, the function $G(\beta)$ was calculated for $0\le\beta\le 2$
and the result
\[
\mid G(\beta)\mid < 10^{-7}
\]
was obtained numerically. 


\end{document}